# On Logical Extension of Algebraic Division

Mohammed Abubakr
Microsoft India (R&D) Pvt. Ltd,
Gachibowli, Hyderabad, INDIA
moabubak@microsoft.com

*Abstract-* **Basic arithmetic is the cornerstone of mathematics and computer sciences. In arithmetic, 'division by zero' is an undefined operation and any attempt at extending logic for algebraic division to incorporate division by zero has resulted in paradoxes and fallacies. However, there is no proven theorem or mathematical logic that suggests that, defining logic for division by zero would result in break-down of theory. Basing on this motivation, in this paper, we attempt at logically defining a solution for 'division by zero' problem.**

## I. INTRODUCTION

Basic arithmetic is the cornerstone of mathematics and computer sciences and popularly considered as well-understood. One operation among basic arithmetic is division of numbers. Division operation is defined as $x/y = z \leftrightarrow [(y \neq 0 \rightarrow x = y.z) \& (y = 0 \rightarrow z \text{ is undefined})]$. It is strange to note that division by zero is assumed as an undefined operation in basic arithmetic. Not surprisingly, this 'division by zero' problem has been one of the oldest open problems in mathematical logic. It is matter of some debate on who first studied this problem, for more details see [1]. In 628 A.D, an Indian mathematician Brahmagupta defined division by zero results in no quotient. Subsequently in 1152 A.D., another Indian mathematician Bhaskara defined division by zero as infinity. Since then, 'division by zero' problem has been extensively studied by several renowned mathematicians and physicists. A noted few are John Wallis, Isaac Newton, John Craig, Martin Ohm, Wolfgang Bolyai de Bolya, De Morgan, William Walton, Rudolf Lipchitz, etc. A good account on history of 'division by zero' problem and related references can be found in [2]. Subsequently during $19^{th}$ century, a consensus was reached to treat division by zero as a non-allowed operation in basic arithmetic.

However, the question still remains open. Is there a possibility of arriving at a mathematical logic that allows division by zero that doesn't result in paradoxes? There is no proven mathematical or logical theorem that states that, allowing division by zero would result in some sort of break-down of mathematical logic. Several logicians have ignored the division by zero as absurd by quoting the variant of the following example. Consider there are 10 apples that have to be divided among 2 people. How many apples will each person get? The answer is 10/2 = 5 apples each. However, if the question is asked, how many apples does each person get if it has to be divided among '0' people? Such question doesn't have any physical meaning as apples can be divided only when there are non-zero people. However, as we point out, such logic cannot be considered as reason for treating division by zero as undefined logic. Consider the following revised question. How many apples does each person get if 10 apples are divided among '2i' people, where 'i' is an imaginary number? Even though, logically '2i' people doesn't make sense, mathematically the answer is well defined and equal to 10/2i = -5i. Hence, we note for a mathematical logic to be true, it necessarily need not have a verbal/logical/physical meaning.

Also, our motivation of working on 'division by zero' problem is because of natural occurrence of division by zero at spacetime singularities within black holes [8-9] and at the big bang [6-7]. Such unavoidable natural occurrence of division by zero conveys two possibilities (i) laws of theoretical physics are incorrect or (ii) there exists some solution to 'division by zero' problem that is unknown to us. We would like to consider the later possibility i.e. there exist some solution to division by zero problem that would help us understanding natural occurrence of division by zero in our physical theories.

This paper is organized in following order. In Section II, in order to define division by zero, we define a new axiom that relates to conservation of information in mathematical expressions. In Section III, division by zero is defined. We show that such division by zero gives rise to new numbers, which we refer as 'Calpanic Numbers'. In section IV, we describe the algebra of numbers arising out of division by zero operation. In Section V, we briefly go over the polynomial equations whose solutions are Calpanic numbers. In section VI, we discuss the properties of Calpanic number matrices. Conditions for eliminating singularities are discussed in Section VII. In the end, we provide a critique and formal conclusion with a optimistic note.

## II. LOGIC, ZERO AND ROLE OF OBSERVER

Logical axioms of arithmetic vary depending on the domain of the numbers i.e. axioms of arithmetic for natural numbers (integers greater than 0) are different from axioms of arithmetic for real numbers. A good treatment of logical axioms of arithmetic operations such as addition and subtraction can be found in [3, 5].

In [3], Suppes describes five elementary techniques that have been adopted at arriving at 'division by zero' problem and various constraints associated with it. The technique adopted in this paper is on the lines of forth technique suggested by Suppes. As per the forth technique suggested by Suppes in

[3], a real number when divided by zero gives rise to non-real number. This technique is analogous to creation of imaginary numbers. Before the advent of imaginary numbers, square-root of a negative number was an undefined arithmetic operation. It was only after assuming the square root of negative number gives rise to imaginary number, we have been able to develop elegant theory of complex numbers. As we now understand, imaginary numbers are not merely a theoretical trick but intrinsically related to core mathematics [12-13].

Before we make an attempt to solve 'division by zero' problem, we briefly discuss some fundamental properties related to 'zero'. In traditional mathematical logic, it is assumed that there is no difference between positive zero and a negative zero. Based on this assumption, it is believed that addition of real numbers is commutative. Addition of two real numbers A, B is commutative i.e. A+B = B+A. Consider the situation, where B = −A

$$A+B = A − A = +0 \quad (A − A = 0; \forall A \in \mathbb{R}^+)$$
$$B+A = −A+ A = (−1)(A − A) = −0$$
$$A+B = +0$$
$$B+A = −0$$
$$(A+B = B+A) \text{ if and only if } +0 = −0$$

As we saw above, addition of real numbers is commutative if and only if positive zero is equal to negative zero. The concept of positive and negative zero is also used in calculus for defining the limits of the real functions [15]. Consider the function $f(x) = 1/x$. At x=0, we define the limits of this function as

$$\lim_{x \to 0^+} \frac{1}{x} = +\infty$$
$$\lim_{x \to 0^-} \frac{1}{x} = -\infty$$

The value of the function f(x) for limit 'x' tends to positive zero is different compared to limit 'x' tends to negative zero. However, at 'x' equals to exact zero, the function f(x) is undefined as it leads to division by zero. The behavior of the function f(x) is important to us, as at x=0, it relates to 'division by zero' problem. f(x) for limit tending to positive zero is different than negative zero indicates the difference between positive and negative zero. Had both positive and negative zero been intrinsically equal, then f(x) would have had the same value for limit tending to positive and negative zero. This gives us a hint that, if 'division by zero' has to be solved, it needs to address the difference between positive and negative zero.

On a side note, almost all computers and programming languages support positive and negative zero. This is referred as signed zero and is defined by IEEE 754 floating point standard. A good account on signed zero can be found in [14]. Hence, if a new mathematical theory threats positive zero different from negative zero, computers and programming languages are in a position to support such change.

We now look at some mathematical properties of 'zero'. Consider the below mathematical equations

$$A + A = 2 \times A; \quad A \in \mathbb{R} \quad (1)$$
$$A + A = A; \quad A \in \mathbb{R} \quad (2)$$

All real numbers satisfy eqn. 1, whereas eqn. 2 is satisfied by only zero. In order words, eqn. 2 is a unique property of real number zero. Eqn. (1) and (2) also lead to another property of 'zero'

$$k \times A = A; \quad \forall k \in \mathbb{R} \quad (3)$$

Eqn. 3 is satisfied only if A=0 and +0 = −0, hence this property is again unique to 'zero' and not satisfied by any other real number. While eqn. (2) and (3) are mathematically correct, an observer performing mathematical calculations can abuse these properties of zero. The following example would highlight how these properties can be abused. Consider a statement, "there are seven mangoes". This can be written mathematically as

$$\text{Mangoes} = 7; \quad (4)$$

However, an observer can write the same statement as eqn. 5 and argue that both eqn. 4 and 5 are numerically equal.

$$\text{Mangoes} = 7 + (0 \times \text{Apples}) \quad (5)$$

Expression (4) and (5) conveys that there are seven mangoes; however eqn. (5) contains more information than specified in the statement 'there are seven mangoes' and the observer who wrote eqn. (5) has abused the properties of zero. This abuse of properties of zero doesn't result in any numerical discrepancy when the calculations are performed in real-number domain, however inserts junk information in the mathematical equation. Insertion of junk information by abusing numerical properties of zero doesn't affect the real number arithmetic, however poses a serious logic challenge for arriving at a solution for 'division by zero' problem. Consider the following situation. Let some non-real quantity U be described as follows

$$U = 1/0 \quad (6)$$

An observer can use the real number properties of zero and modify U as

$$U = 1/0 = 1/0 \times (1/5) = 5/0$$
Similarly, $\quad U = 1/0 = r/0 \quad \forall r \in \mathbb{R}-\{0\} \quad (7)$

Therefore, if we define some non-real number U = 1/0 and assume that properties of real number 'zero' hold good in the new domain, then U should also be equal to r/0 for ∀ r ∈ ℝ-{0}. The negation of this logic is that, if some non-real quantity U = 1/0 has to be unique and not equal to r/0 for ∀ r ∈ ℝ -{1}, then one of the necessary conditions is that properties of zero doesn't hold good in new domain.

From the above two examples, in the new domain of numbers where division by zero is allowed, the number zero must have following properties

(i) Negative zero is not equal to positive zero
(ii) An observer cannot abuse the properties of zero given in Eqn. 2 and 3.

We now define an axiom that would address the properties of 'zero' in the new domain.

*Axiom 1: An observer or a machine performing mathematical calculations must not create or destroy information using zero.*

We refer the above axiom as 'conservation of information'. This axiom prohibits abuse of numerical properties of zero by the observer. Using the above axiom, we obtain the following properties of zero. Note that, these properties are only valid for new domain of numbers where division by zero is allowed.

(i) If A = 0, then an observer cannot write A = 0 + 0. Similarly, if A = 0 + 0, then observer cannot write it as A = 0. However, the observer can write A = 0 + 0 as 2 × 0.
(ii) If A = 0, then an observer cannot write A = 0 × 0. Similarly, if A = 0 × 0, then observer cannot write it as A = 0. We refer, 0 × 0 as second order zero and 0 as first order zero.
(iii) Multiplication of a real number with zero is prohibited as it creates or destroys information. This means that +1×0 is not same as −1×0.

Consider the following mathematical expressions

$$x = z^2 \qquad (8)$$
$$y = z - 1 \qquad (9)$$

At z = 0, x = $0^2$, i.e. z =0, x is a second order zero. Similarly, at z = 1, y is first order zero. Given that, conservation of information axiom states that 'observer' cannot kill the information, a second order zero must not be converted into first order zero in a mathematical expression i.e. if x = 0 × 0, then the expression must not be converted to x = 0.

In order to simply things, we follow the below notation. During all the step-by-steps proofs of mathematical statements, information is conserved. From here on, real number calculations\measurements are performed only within brackets [ ] $_{real}$. For example, consider eqn. 8

$$x = 0^2 \qquad \text{for z = 0;}$$
$$[x]_{real} = 0$$

In some crude sense, doing a real calculation can be considered similar to 'wave function collapse' in quantum mechanics. Once the observer measures the real state of the observable, the wave function collapses to a real value [10-11]. Similarly, once the real calculation is performed, information collapses to a value.
We shall discuss the other consequences of 'conservation of information' axiom in next section.

III. DIVISION BY ZERO

We now define division by zero as follows:

*Definition 1*: Let 'क' be some non-real number such that

$$क = \frac{1}{0} ; \ 0 = \frac{1}{क} ; \ क \times 0 = 0 \times क = 1$$

*Definition 2*: The non-real number 'क' satisfies the following

$$z \times क = zक = कz ; \ \forall z \in \mathbb{R}$$

We call the numbers containing "क" as a Calpanic numbers[1]. We now define a combination of Real and Calpanic number as follows

$$B = z_0 + z_1 क^1 ; \ z_0, z_1 \in \mathbb{R}$$

For simplicity, from hereon we refer Real + Calpanic numbers as Calpanic numbers itself and we refer its domain of real and Calpanic numbers as calpanic domain. Before we proceed to derive theorems relating to Calpanic numbers, we formally defines the rules for calpanic numbers. These rules are derived from axiom of 'conservation of information'.

Rule 1: Multiplication of a real number with zero destroys the information, therefore in Calpanic domain, zero must not be multiplied with real numbers. Where ever possible, definition 1 must be applied to eliminate '0' and 'क'.

An important consequence of rule 1 is +0 ≠ − 0, since 0 = 1×0 and −0 = −1×0. Positive and negative zero are equal only if real number multiplication with zero is allowed. Another consequence of this rule is that, क ×−0 = −1.

Rule 2: The order of the zero must be preserved. This follows from the Rule 1 which prohibits multiplication of real numbers with zero. This means first order zero is not equal to second order zero.

---

[1] The symbol 'क' is a Sanskrit alphabet and it is pronouced as 'ka'. The word 'Calpanic' means 'un-real' in Sanskrit.

Rule 3: In Calpanic domain, following approach must be adopted while subtracting two numbers B, A in order to conserve information. First consider the special case, B = A

$$B - A = A - A$$
$$= A(1-1)$$
$$= (A \times 0)$$

Ex: $2 - 2 = 2(1-1)$
$$= 2 \times 0$$

This can be generalized as

$$A - A = A \times 0; \forall A \in \mathbb{R}$$

Note that rule 1 prohibits multiplication of real number with zero. Now, let us consider the general case where $B = A + C$ & $C \neq 0$

$$B - A = (A + C) - A$$
$$= C + A(1-1)$$
$$= C + (A \times 0)$$

Note that, $[A - A]_{real} = [A \times 0]_{real} = 0$.
Also, the observer cannot introduce information in the mathematical expression by adopting following approach.

Let $B = A + C$, where $C \neq 0$
$$B - A = (A + C) - A$$
$$= C + A(k - k)/k; \text{ where } k \neq 0$$
$$= C + A(0)/k$$

Here, the observer is killing the information by writing '$k - k = 0$' instead of '$k - k = k \times 0$'.

Rule 4: For some Calpanic number $P = x_1 + x_2 + \ldots + x_n$ such that $x_1, x_2, \ldots, x_n = 0$, then $P = n \times 0$. Similarly, if $x_1 = 0$ and $x_2, x_3 \ldots x_n \neq 0$, then

$$P = 0 + x_2 + x_3 \ldots + x_n$$

Any violation of above rules would result in paradoxes. Definition 1-2 and rules 1-4 give rise to a rich Calpanic number theory. We now derive theorems related to Calpanic Numbers.

*Theorem 1: Addition is not commutative in Calpanic domain.*
*Proof:* We prove this theorem using proof by contradiction. Let A = '1' and B = '- 1'.

$$A + B = 1 - 1 = +0$$
$$B + A = -1 + 1 = (-1)(1 - 1)$$
$$= (-1)(0)$$
$$= -0$$

Above, we have made an explicit assumption that $1 - 1 = + 0$. Upon multiplying with – 1 on both sides, we would get $-1 + 1 = -0$. However, we can also assume $1 - 1 = -0$ and $-1 + 1 = +0$. Irrespective of our assumption, we would $(A+B) = -(B+A)$. In Calpanic domain $+0 \neq -0$ from rule 1, hence there are some values of A, B for which $A+B = B + A$ is not satisfied. Hence, addition is not commutative in Calpanic domain.

Traditionally, addition has been commutative and several of our mathematical conventions are dependent on it. However, since addition is not commutative in Calpanic domain, we address some of the convention changes here. It is the matter of convention to choose whether $1-1 = +0$ or $-0$. From here on we abide to the following convention.

$$1 - 1 = + 0 \quad (10)$$
$$-1 + 1 = - 0 \quad (11)$$

Note that, this convention doesn't change Rule 3. Following rule still holds good.

$$A - A = A \times 0; \forall A \in \mathbb{R}$$

Consider the special case, A = 0

$$0 - 0 = 0(1 - 1) = 0 \times 0$$
Hence, $\quad 0 - 0 = 0 \times 0 \quad (12)$

*Theorem 3: For a Calpanic number A, A – A not necessarily equal to 0.*
*Proof:* We prove this theorem using proof by contradiction. Consider the special case where A = 0. From eqn. 12,

$$0 - 0 = 0(1 - 1) = 0 \times 0 \neq 0.$$

Hence, for a Calpanic number A, A - A necessarily need not be equal to zero.

As per traditional logic of subtraction, for a real number A, A - A = 0. However, in Calpanic domain, such traditional logic is not applicable. Consider the case of A = B = -1

$$A - B = -1 + 1 = - 0$$

Therefore, we get a condition that, if A = A then A – A is equal to 0 if $+ 0 = - 0$. However, in Calpanic domain positive zero is not equal to negative zero.

*Corollary 2.1*: For four Calpanic numbers A, B, C and D, such that $A + B = C + D$, then $A + B - D$ need not be equal to C.
*Proof:* $A + B = C + D$
Subtracting D on both sides
$$A + B - D = C + D - D$$
From theorem 2, D – D is not equal to 0. Therefore, if $A + B = C + D$ then $A + B - D \neq C$.

*Corollary 2.2*: क – क = 1
*Proof:* क – क = क (1 – 1)
$$= क (0) \quad \text{(from eqn. 10)}$$
$$= 1 \quad (क0 = 1 \text{ from Def. 1})$$
Hence, क – क = 1

This property of Calpanic numbers is most surprising and useful in eliminating क.

*Corollary 2.3*: $-$ क $+$ क $= -1$
Proof:  $\quad -$ क $+$ क $= (-1)($ क $-$ क $)$
$\qquad\qquad\qquad = (-1)(+1)$
$\qquad\qquad\qquad = -1$

*Theorem 3: Addition of Calpanic numbers is not necessarily associative i.e. (A+B) + C ≠ A + (B+C) for some values of A, B and C.*
Proof: We prove this theorem by contradiction. Let A = क, B = $-$ क and C = क.

$(A+B) + C = ($ क $-$ क $) +$ क
$\qquad\qquad\;\; = 1 +$ क $\qquad$ (from corollary 2.2)
$A + (B+C) =$ क $+ (-$ क $+$ क $)$
$\qquad\qquad\;\; =$ क $-$ क $(1 - 1)$
$\qquad\qquad\;\; =$ क $- 1 \qquad$ (from corollary 2.3)

From above, (A+B) +C = 1 + क, however A + (B+C) is equal to क - 1 i.e. *(A+B) + C ≠ A + (B+C)*. Hence, addition of Calpanic numbers is not necessarily associative. Non-associative nature of addition poses a challenge during calculations if a proper convention is not followed. From here on, in all our calculations, we perform addition from 'left to right'.

*Theorem 4*: LOG (0) = –LOG (क)
Proof:  From definition 1, $0 = 1/$ क $\Rightarrow$ क $= 0^{-1}$
Taking LOG on LHS and RHS

$\qquad$ LOG (क) = LOG($0^{-1}$)
$\qquad$ LOG (क) = –LOG(0)
$\Rightarrow \qquad$ LOG (0) = –LOG(क)

*Corollary 4.1*: क$^{+0}$ = क$^{-0}$ = 1
*Proof*:
$LOG\ (0) = -LOG\ ($ क $)$
$0 = e^{-LOG($ क $)}$
क $= e^{(-LOG(0)} = e^{LOG($ क $)}$
क $\times 0 = e^{LOG($ क $)} e^{-LOG($ क $)}$
$\qquad\; = e^{LOG($ क $) - LOG($ क $)}$
$\qquad\; = e^{(1-1)LOG($ क $)}$
$\qquad\; = e^{(0)LOG($ क $)}$
$\qquad\; = e^{LOG\ \text{क}^0}$
क $\times 0 = e^{LOG($ क$^0)}$
$e^{LOG($ क$^0)} = 1 \qquad\qquad \because$ क $\times 0 = 1$
$e^{LOG($ क$^0)} = e^0$
$\Rightarrow LOG($ क$^0) = 0$
$\Rightarrow$ क$^0 = 1 \qquad\qquad \because LOG\ (1) = 0$
$0 \times$ क $= e^{-LOG($ क $)} e^{LOG($ क $)}$
$\qquad\; = e^{-LOG($ क $) + LOG($ क $)}$
$\qquad\; = e^{(-1+1)LOG($ क $)}$
$\qquad\; = e^{(-0)LOG($ क $)}$
$\qquad\; = e^{LOG\ \text{क}^{-0}}$
$0 \times$ क $= e^{LOG($ क$^{-0})}$
$e^{LOG($ क$^{-0})} = 1 \qquad\qquad \because 0 \times$ क $= 1$
$e^{LOG($ क$^{-0})} = e^0$
$\Rightarrow LOG($ क$^{-0}) = 0$
$\Rightarrow$ क$^{-0} = 1 \qquad\qquad \because LOG\ (1) = 0$

$\therefore$ क$^{+0} =$ क$^{-0} = 1$

IV. CALPANIC NUMBER ALGEBRA

Using theorem 1, 2 and 3 we now discuss the algebra of Calpanic numbers. Consider the Calpanic number as given in definition 3

$$B = z_0 + z_1 \text{क}^1; \quad z_0, z_1 \in \mathbb{R} \qquad (13)$$

In eqn. 13, $z_0$ is called real component and $z_1$ is Calpanic component of B. Note that, if $z_1$ is equal to zero, then क in (13) is eliminated. We call क$^n$ as $n^{th}$ order singularity and $0^m$ as $m^{th}$ order zero. A generic mixed number with $n^{th}$ order singularity and $m^{th}$ order zero can be written as

$$B_{MN} = \sum_{n=0}^{N} Z_n \text{क}^n + \sum_{m=1}^{M} W_m 0^m; \quad \forall Z_n, W_m \in C\ \&\ Z_0, Z_N, W_M \neq 0;$$

$$B_{0N} = \sum_{n=0}^{N} Z_n \text{क}^n \qquad (14)$$

We use simple Calpanic numbers in all our examples for ease of calculations, similar logic can be extended for Calpanic numbers of $n^{th}$ order singularity and $m^{th}$ order zeros.

Consider the addition of two Calpanic numbers $B_1$ and $B_2$ given by $a_0 + a_1$ क and $b_0 + b_1$ क respectively, then

$B_1 + B_2 = a_0 + a_1$ क $+ b_0 + b_1$ क
$\qquad\qquad = (a_0 + b_0) + (a_1 + b_1)$ क $\qquad (15)$

Similarly,

$B_1 - B_2 = (a_0 - b_0) + (a_1 - b_1)$ क

$B_1 \times B_2 = (a_0 + a_1$ क $) \times (b_0 + b_1$ क $)$
$\qquad\qquad = a_0 b_0 + (a_0 b_1 + b_0 a_1)$ क $+ a_1 b_1$ क$^2$

Similarly, if $B_1 = B_2$

$B_1 + B_1 = 2(a_0 + a_1$ क $) = 2B_1$

$B_1 - B_1 = (a_0 - a_0) + (a_1 - a_1)$क
$\quad\quad\quad\quad = a_0 \times 0 + a_1 ($क$-$क$)$
$\quad\quad\quad\quad = a_0 \times 0 + a_1$

if $a_1 \neq 0$ then $B_1 - B_1 \neq 0$

$[B_1 - B_1]_{real} = [a_0 \times 0 + a_1]_{real} = a_1$

$B_1 \times B_1 = a_0^2 + 2a_0 a_1$ क $+ a_1^2$ क$^2$

Other properties of Calpanic numbers include

1. Multiplication of Calpanic numbers is not necessarily commutative.
2. [AB – BA] is not necessarily equal to 0, where A, B are two Calpanic numbers.

Following examples demonstrate how calculations are performed in Calpanic domain

- $x/0 = x$ क ; $\forall x \in \mathbb{R}$
  Ex: $5/0 = 5(1/0) = 5$ क
  Note: As per axiom of conservation of information, an observer cannot convert the demoninatior '0' to '0×k'. Also, the observer must not multiply the real number with zero.

- $X - Y = [X - Y]_{real} + (0 \times Y)$; $\forall X, Y \in \mathbb{R}$ & $X > Y$
  Ex: Let X=10; Y=7
  $X - Y = [10–7]_{real} + (0 \times 7)$
  $\quad\quad = 3 + (0 \times 7)$
  Note: $[X-Y]_{real} = 3$
  This notation might be little confusing given that we are used to traditional subtraction in a different way. In Calpanic Number Algebra, 'zero' plays an important role and can't be ignored.

- $X$ क $- Y$ क $=$ क $[X–Y]_{real} + Y$ ; $\forall X, Y \in \mathbb{R}$ & $X > Y$
  Ex: Let $X = 10$; $Y = 7$
  $X$ क $- Y$ क $= 10$ क $- 7$ क $= 3$ क $+ 7$ (क$-$क)
  $\quad\quad\quad\quad = 3$ क $+ 7$ ; (since क$-$क$= 1$)
  As seen in this example, 10 क -7 क is equal to '3 क + 7' but not '3 क'.

## V. POLYNOMIAL EQUATIONS WITH CALPANIC NUMBERS

Consider the polynomial equations such as below

$$x - x = k \quad ; k \in \mathbb{R} - \{0\} \quad (16)$$
$$x^2 - x^2 = k_1 \text{क} + k_0 \quad (17)$$

In complex or real number domain, eqn. 16 has no meaning for $k \neq 0$. In case of Eqn. 17, the $x^2$ term can be cancelled without hesitation in complex or real number domain, resulting in $k_1 x + k_0 = 0$. However, in Calpanic number domain, these equations (i.e. Eqn. 16 and Eqn. 17) needs to be solved recognizing the fact that x-x $\neq$ 0.

As per Eqn. 10, $x - x = k$

To solve the polynomial equations, we have to identify some value of 'x' for which L.H.S = R.H.S. Given that if A+B = C+D, then A+B-D might not be equal to C, we cannot shift a variables or constants from left hand side to right hand side or vice versa. Let us consider, there is some 'x' such that x = $x_1$ क that satisfies eqn. 10.

$\Rightarrow x - x = x_1($क$-$क$) = k$
$\quad\Rightarrow x_1 = k$

$\Rightarrow x_1 = c$. Note that, we have only found solutions for $x = x_1$ क, however there is also a possibility that some higher order x with nth order singularity and mth order zero would satisfy the equation x -x = c. Also, a word of caution here, the equation x-x = c is not equal to x = x+c (from theorem 2).

A general result is that, $x = k$ क are solutions to polynomial equations of below type

$$x^n - x^n = k\, x^{n-1} \;; \quad n \in N$$

At this point of time, we haven't been able to find a procedural technique for solving polynomial equations involving Calpanic numbers. Identifying general properties of polynomial equations involving Calpanic numbers is one of our future goals.

## VI. MATRICES WITH CALPANIC NUMBERS

We have observed break-down of traditional matrix concepts when Calpanic numbers are used in matrices. Multiplying a Calpanic number matrix with the traditional identity matrix doesn't produce the same Calpanic Number matrix. Consider the case for a generic 2×2 matrix,

$$A = \begin{bmatrix} a_{11} + b_{11}\text{क} & a_{12} + b_{12}\text{क} \\ a_{21} + b_{21}\text{क} & a_{22} + b_{22}\text{क} \end{bmatrix} ; I = \begin{bmatrix} 1 & 0 \\ 0 & 1 \end{bmatrix}$$

$$a_{mn}, b_{mn} \in \mathbb{R}$$

$$AI = \begin{bmatrix} (a_{11} + b_{11}\text{क}) + b_{12} & b_{11} + (a_{12} + b_{12}\text{क}) \\ (a_{21} + b_{21}\text{क}) + b_{22} & b_{21} + (a_{22} + b_{22}\text{क}) \end{bmatrix} \neq A$$

$$IA = \begin{bmatrix} (a_{11} + b_{11}\text{क}) + b_{21} & b_{22} + (a_{12} + b_{12}\text{क}) \\ (a_{21} + b_{21}\text{क}) + b_{11} & b_{12} + (a_{22} + b_{22}\text{क}) \end{bmatrix} \neq A$$

If A is a real number matrix i.e. $b_{11} = b_{22} = b_{21} = b_{12} = 0$, then $AI = IA = A$ holds good. This can be verified by substituting $b_{11} = b_{22} = b_{21} = b_{12} = 0$ in above matrices.

A Calpanic number matrix A satisfying the condition $b_{11} = b_{22}$ and $b_{21} = b_{12}$, would satisfy the following

$$IA = AI$$

However, $b_{11} = b_{22}$ and $b_{21} = b_{12}$ is not a sufficient condition to prove $IA$ or $AI = A$. Hence, I is the special case for real-number matrix. During our analysis, we have found that determinant of AI is not necessarily equal product of determinant A and determinant I. Hence, in general for any two square matrices A and B of same order, det(AB) is not necessarily equal to det(A) × det(B).

## VII. Elimination of Singularities

In this section, we study the conditions for addition of two Calpanic numbers which eliminates singularities. Consider two simple Calpanic numbers $B_1$ and $B_2$ given by $a_0 + a_1 क$ and $b_0 + b_1 क$ respectively, such that $a_0, b_0, a_1, b_1 \in C-\{0\}$.

$$B_1 + B_2 = (a_0 + b_0) + (a_1 + b_1)क$$

If $b_1 = -a_1$, then first order singularity would eliminated.

$$\begin{aligned} B_1 + B_2 &= (a_0 + b_0) + (a_1 - a_1)क \\ &= (a_0 + b_0) + a_1(क - क) \\ &= (a_0 + b_0) + a_1 \end{aligned}$$

Note that, if $b_1 = -a_1$, it also implies $a_1 = -b_1$. Substituting $a_1 = -b_1$,

$$\begin{aligned} B_1 + B_2 &= (a_0 + b_0) + (-b_1 + b_1)क \\ &= (a_0 + b_0) + b_1(-क + क) \\ &= (a_0 + b_0) + b_1(-1)(क - क) \\ &= (a_0 + b_0) - b_1 \\ &= (a_0 + b_0) + a_1 \end{aligned}$$

Hence, if $b_1 = -a_1$ or $a_1 = -b_1$, the value of $B_1 + B_2$ remains unchanged. If $(a_0 + b_0) = a_1$, then $[B_1 + B_2]_{real}$ is equal to '0'.

Similarly, for addition of Calpanic numbers containing second order singularity, conditions can be obtained for which singularities are eliminated. If the singularity of the black-hole can be described using Calpanic numbers, then additive properties of Calpanic numbers can potentially throws light on black hole dynamics and conditions for which singularities can be eliminated. However, this isn't to say that Calpanic numbers does indeed find application in black-hole dynamics. Further research needs to be carried out on to establish whether Calpanic numbers can be used to describe the black-hole singularity and dynamics or not.

## VIII. Calpanic Numbers – A critique

In previous sections of this paper, we have formulated the new theory of Calpanic numbers. An obvious question at this moment is, is this theory worth formulating given that fact that, it suggests abandoning commutatively and associativity of numbers. Moreover, the theory bizarrely suggests negative zero is not equal to positive zero. Hence, it is natural ask, whether this Calpanic number theory adds any value of mathematics. However, there is another side of the coin. Real numbers are merely a subset of Calpanic numbers. Hence, the logic associated with real numbers is special case of Calpanic number logic. For a given application, if all the associated Calpanic numbers are purely real numbers then, addition is associative and commutative.

We would also like to highlight that, any mathematical theory is considered valuable if only it finds its applications. Historically, complex numbers were seen with distrust and misbelief until applications of it were found. Hence, Calpanic numbers are valuable if only they find physical applications. Currently, we are working on potential applications of Calpanic theory in applied fields such as cryptography. We also foresee applications of Calpanic numbers in representation of curled dimensions, black-hole dynamics, etc.

## IX. Conclusions

In this paper, we have presented a novel concept by considered division by zero gives rises to non-real numbers called as Calpanic number. The approach adopted in this paper is similar to approach adopted for creation of imaginary numbers by considering square-root of minus one is equal to imaginary number 'i'. By considering one divided by zero as a Calpanic number represented by symbol 'क', we explored the possibility of extending division algebra. Subsequently, we derived Calpanic number algebra and theory. The Calpanic number theory presented in this paper is still under evolutionary state and has huge scope for further development.

## Acknowledgements

I would like to thank Sumiya Faruq for verifying the proofs and for numerous suggestions that improved this paper. The original idea of this paper was first conceived in 2005 and was abandoned in 2007. However, it was during International Congress of Mathematicians (ICM 2010) held at Hyderabad, during the discussions with mathematicians, I was encouraged to publish the idea. I would like thank Microsoft Research India for sponsoring me to attend ICM 2010 and the various mathematicians who encouraged me to revive this idea.

## References

[1] Boyer C. B, "An Early Reference to Division by Zero", The American Mathematical Monthly, Vol. 50, No. 8, pp. 487-491, (1943).
[2] Romig H. G, Early History of Division by zero, The American Mathematical Monthly, Vol. 31, No. 8, pp. 387-389, (1924).
[3] Patrick Suppes, *Introduction to Logic*, Litton Educational publishing, pp. 128-176, (1957).
[4] Shawn Hedman, *A first course in Logic: An Introduction to Model Theory, Proof Theory, Computability and Complexity*, Oxford University Press, 2004.
[5] Alfred Tarski, Introduction to Logic and to the MethodoLOGy of Deductive Sciences, Oxford University Press, 1994.
[6] Steven Weinberg, Cosmology, Oxford University Press, 2008
[7] John F. Hawley and Katherine A. Holcomb, Foundations of Modern CosmoLOGy, Oxford University Press, 2005
[8] Pankaj S. Joshi, Gravitational Collapse and Spacetime Singularities, Cambridge University Press, 2007
[9] Valeri P. Frolov and Igor D. Novikov, *Black hole physics: Basic Concepts and New Developments*, Kluwer Academic Publishers, 1998
[10] Orly Alter and Yoshihisa Yamamoto, *Quantum Measurement of a Single System*, John Wiley & Sons, 2001.
[11] S. Mayburov, Quantum Information and Wave function Collapse, arXiv:0807.2768v1 [quant-ph], 2008
[12] Peter Borwein, Stephen Choi, Brendan Rooney and Andrea Weirathmueller, *The Riemann Hypothesis: A Resource for the Afficionado and Virtuoso Alike*, Springer, 2007.
[13] H. M. Edwards, *Riemann's Zeta Function*, Dover Publications, 1974
[14] David Goldberg, What every computer scientist should know about floating-point arithmetic, Vol. 23, Issue 1, ACM Computing Surveys, 1991.
[15] G.H. Hardy, A course in pure mathematics, Cambridge University Press, 1921.